\documentclass[a4paper,11pt]{article}
\usepackage{pos}
\usepackage{orcidlink}
\usepackage{subfig}
\usepackage{diagbox}
\usepackage{makecell, multirow}
\usepackage{pgf}
\usepackage{soul}
\usepackage{graphicx, color}
\usepackage{rotating}
\usepackage{braket}
\usepackage{mathtools}
\allowdisplaybreaks

\definecolor{link_blue}{RGB}{51,102,204}
\hypersetup{%
pdftitle = {title},
pdfsubject = {},
pdfkeywords = {},
colorlinks = {true},
filecolor = {black},
linkcolor = {link_blue},
menucolor = {black},
citecolor = {link_blue},
urlcolor = {link_blue},
}{}

\title{Optimization of the HHL Algorithm}

\author*{Dhruv Sood}
\author{Nilmani Mathur}
\author{Vikram Tripathi}

\affiliation{Department of Theoretical Physics, Tata Institute of Fundamental Research, Homi Bhabha Road, \\
  Mumbai 400005, India}

\emailAdd{dhruv.sood@tifr.res.in}
\emailAdd{nilmani@theory.tifr.res.in}
\emailAdd{vtripathi@theory.tifr.res.in}

\abstract{The Harrow–Hassidim–Lloyd (HHL) algorithm is a quantum algorithm for solving systems of linear equations that, in principle, offers an exponential improvement in scaling with the system size compared to classical approaches. In this work, we investigate the practical implementation and optimisation of the HHL algorithm with a focus on improving its performance on near-term quantum simulators. After outlining the algorithm, we examine two optimisation strategies aimed at improving fidelity and scalability: Suzuki–Trotter decomposition of the Hamiltonian evolution operator and a block-encoding approach that embeds the problem matrix into a larger unitary operator. The performance of these methods is evaluated through simulations on matrices with varying sparsity, including diagonal, tridiagonal, moderately dense, and fully dense cases. Our results show that while HHL achieves near-ideal fidelity for highly structured matrices, performance degrades as sparsity decreases due to the increasing cost of Hamiltonian simulation and reduced post-selection probability due to higher condition number. Block encoding is found to provide improved fidelity for moderately dense matrices, whereas Trotterisation offers a qubit-efficient approach for sparse systems. These results highlight the importance of matrix structure in determining the practical efficiency of HHL and inform future implementations that combine algorithmic optimisation with hardware-aware design.}

\FullConference{
}

\begin{document}
\maketitle

\section{Introduction}
Quantum mechanics emerged in the 1900s to try and explain phenomena at the scales of atoms and revolutionised our understanding of nature, leading to advances across all fields of science and technology. Its success brought with itself the realisation that the classical picture of the world could not hope to describe the subtleties that characterise the physics of very small scales. Indeed, since the world is being proved to be inherently quantum on these scales, it is only natural that any purely classical method of studying it will fail to grasp all aspects of the physical system of interest. Quantum computation offers a solution to this problem, replacing the manipulation of classical bits with those of qubits whose evolution is governed by the laws of quantum mechanics.

Such applications are often quantum algorithms that are able to solve certain problems much faster than the corresponding classical algorithms. Prototypical examples of these  are Shor’s \cite{shor94} algorithm - which allows one to find the prime factors of integers in polynomial time - and Grover’s \cite{grover96} algorithm - which solves the unstructured search problem in $O(\sqrt N)$ time as opposed to the classical $O(N)$. In this work, we will describe results obtained regarding another such quantum algorithm that is able to solve a system of linear equations faster than standard classical algorithms that try to achieve the same. 

Linear Systems of equations are ubiquitous in all fields of science. A more efficient procedure of solving them has the potential to drastically accelerate studies of physical systems. Classically $O(N^{2.81})$ is the best one can currently achieve for matrix multiplication \cite{matmul}, and the solution of $A\mathbf x=\mathbf b$ will obviously be bounded by this. So a quantum algorithm that can achieve an exponential speedup can greatly benefit any computation that can be reduced to such a problem.

The algorithm of interest is the HHL algorithm presented by by Harrow, Hassidim and Lloyd \cite{PhysRevLett.103.150502}. in 2009. It allows one to find an (approximate) solution to the linear system of equations $A\mathbf x=\mathbf b$, given some previous knowledge of the spectrum of $A$ and that the matrix $A$ matrix that is Hermitian, sparse and well-conditioned. The HHL algorithm is able to solve such a problem in time that scales logarithmically in $N$ and polynomially in the condition number $\kappa$ of the problem. The solution is obtained as the amplitudes of a quantum state, so it lends itself naturally to situations where one is not interested in the solution itself, but in some linear function of the same.

The first step in this direction was the HHL algorithm of Harrow, Hassidim and Lloyd \cite{PhysRevLett.103.150502}. It allows one to find an (approximate) solution to the linear system of equations $A\mathbf x=\mathbf b$, given some knowledge of the nature of the spectrum of $A$ and requiring that the matrix $A$ is Hermitian, sparse and well-conditioned. Under reasonable assumptions (efficient access to $A$ and $\ket b$, $A$ is $s$-sparse, the query complexity of the HHL algorithm scales as $\text{poly}(\log N)s^2 \kappa^2/\epsilon,$ where $\kappa$ is the condition number of the problem and $\epsilon$ is a measure of the precision. The exponential speedup in $N$ thus comes at the cost of polynomial overhead in $\kappa$ and $\epsilon$. The solution is obtained as the amplitudes of a quantum state, so it lends itself naturally to situations where one is not interested in the solution itself, but in expectation values of operators in the solution state.

The HHL algorithm has had a significant impact on the state of both current and future prospects of quantum computation and many applications of the same have already been proposed. For eg. Clader et al. \cite{clader13} have proposed its application in speeding up the calculation of electromagnetic scattering cross-sections while Wang et al. \cite{wang17} have put forward using it to estimate the resistances of electrical networks. There is also recent interest in studying it's implementation and it's extensions to matrices with higher condition numbers\cite{calcuttagroup25}.

\section{The HHL Algorithm}

The core idea of the HHL algorithm is to encode the input vector $\mathbf{b}$ into a quantum state $\ket b$ and then use quantum operations to transform this state into the state $\ket x$ that is proportional to the solution $\mathbf{x}=A^{-1}\mathbf b$ for an $N\times N$ matrix $A$. This is achieved by exploiting the completeness of the eigenstates of $A$ to write $\ket b=\sum_{i=0}^{N-1}b_i\ket{v_i}$. Similarly their orthonormality leads to, 
\begin{equation}
  \begin{aligned}
    \ket x&=\frac{A^{-1}\ket b}{||A^{-1}\ket b||_2},  \\
    &=\frac{1}{||A^{-1}\ket b||_2}\sum_{j=0}^{N-1}\frac{\ket{v_k}\bra{v_k}}{\lambda_k}\sum_{k=0}^{N-1}b_j\ket{v_j}, \\
    &=\frac{1}{\sqrt{\sum_{j=0}^{N-1}|b_j|^2/\lambda_j^2}}\sum_{j=0}^{N-1}\frac{b_j}{\lambda_j}\ket{v_j}.
  \end{aligned}
\end{equation}

\begin{figure}[h]
                \centering
                \includegraphics[width=\textwidth]{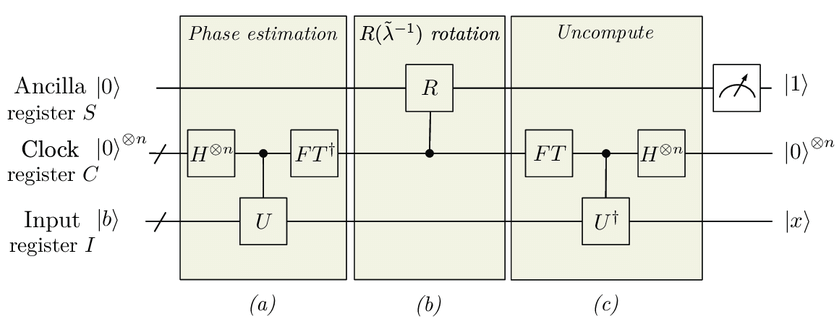}
                \caption{HHL Circuit Diagram }
                \label{fig:blockI}
\end{figure}

This is implemented in two main steps. The first step extracts the eigenvalues of $A$ into a separate register of qubits using Quantum Phase Estimation. Following that, a set of controlled rotations by this register insert the desired factor of the inverse eigenvalues. Figure \ref{fig:blockI} shows a schematic diagram of a circuit that implements this.

The first phase of the circuit, Quantum Phase Estimation, takes a unitary matrix $U$ with eigenstates $\ket{v_j}$ with eigenvalues $\mathrm e^{2\pi i\theta_j}$ and stores the the binary representation of
the phase angle $\theta$ in a qubit state. These qubits carrying the phase information are named “clock qubits". In the HHL algorithm, if $\ket{v_j}$ is an eigenstate of a Hermitian matrix $A$ with eigenvalue $\lambda_j$, by constructing a unitary matrix $U=\mathrm e^{itA}$ with a scale factor
$t$, the state $\ket{v_j}$ becomes an eigenstate of $U$ with eigenvalue $\mathrm e^{it\lambda_j}$ . Therefore, the eigenvalue $\lambda_j$ can be estimated using the QPE algorithm. The second step uses an ancillary qubit and these encoded eigenvalues to insert the desired factor of $1/\lambda_j$ to obtain the solution. Finally, the un-compute step returns the clock qubits to their original state before measurement.

Without circuit optimization, the increase in the system size exponentially increases the gate counts in HHL circuits. Every extra clock qubit
leads to an extra controlled $U^{2^{n_c}-1}$ in the QPE section. As a result, QPE involves $\sum_{j=0}^{n_c-1}2^j=2^{n_c}-1$ applications of $U$. Hence, the primary focus for using this algorithm efficiently is the optimisation in the implemention of QPE, and that is $U$, efficiently.

In general, the qubit states in the HHL circuit evolve as

\begin{equation}
  \begin{aligned}
    \ket 0\ket 0^{\otimes n_c}\ket 0^{\otimes n_d}&\xrightarrow{\text{State Preparation}}\ket 0\ket 0^{\otimes n_c}\ket b,  \\
    &\xrightarrow{\text{QPE}}\sum_{j=0}^{2^{n_d}-1}b_j\ket 0\ket{\lambda_j}\ket{v_j}, \\
    &\xrightarrow{\text{Eigenvalue Inversion}}\sum_{j=0}^{2^{n_d}-1}b_j\Bigg(\sqrt{1-\frac{C^2}{\lambda_j^2}}\ket 0+\frac{C}{\lambda_j}\ket 1\Bigg)\ket{\lambda_j}\ket{v_j}, \\
    &\xrightarrow[\ket a=\ket 1]{\text{Measure and only keep}}C'\ket 1\sum_{j=0}^{2^{n_d}-1}\frac{b_j}{\lambda_j}\ket{\lambda_j}\ket{v_j}, \\
    &\xrightarrow{\text{Inverse QPE}}C'\ket 1\ket 0^{\otimes n_c}\sum_{j=0}^{2^{n_d}-1}b_j\frac{b_j}{\lambda_j}\ket{v_j}=C'\ket 1\ket 0^{\otimes n_c}\ket x.
  \end{aligned}
\end{equation}

Under reasonable assumptions (efficient access to $A$ and $\ket b$, $A$ is $s$-sparse and has condition number $\kappa$), HHL runs in time polynomial in $\log N$, the sparsity and $\kappa$, and the inverse precision $1/\epsilon$. The exponential speedup in $N$ thus comes at the cost of polynomial overhead in $\kappa$ and $\epsilon$.

\begin{figure}[!h]
    \centering
    \begin{minipage}[b]{0.49\textwidth}
        \includegraphics[width=\textwidth]{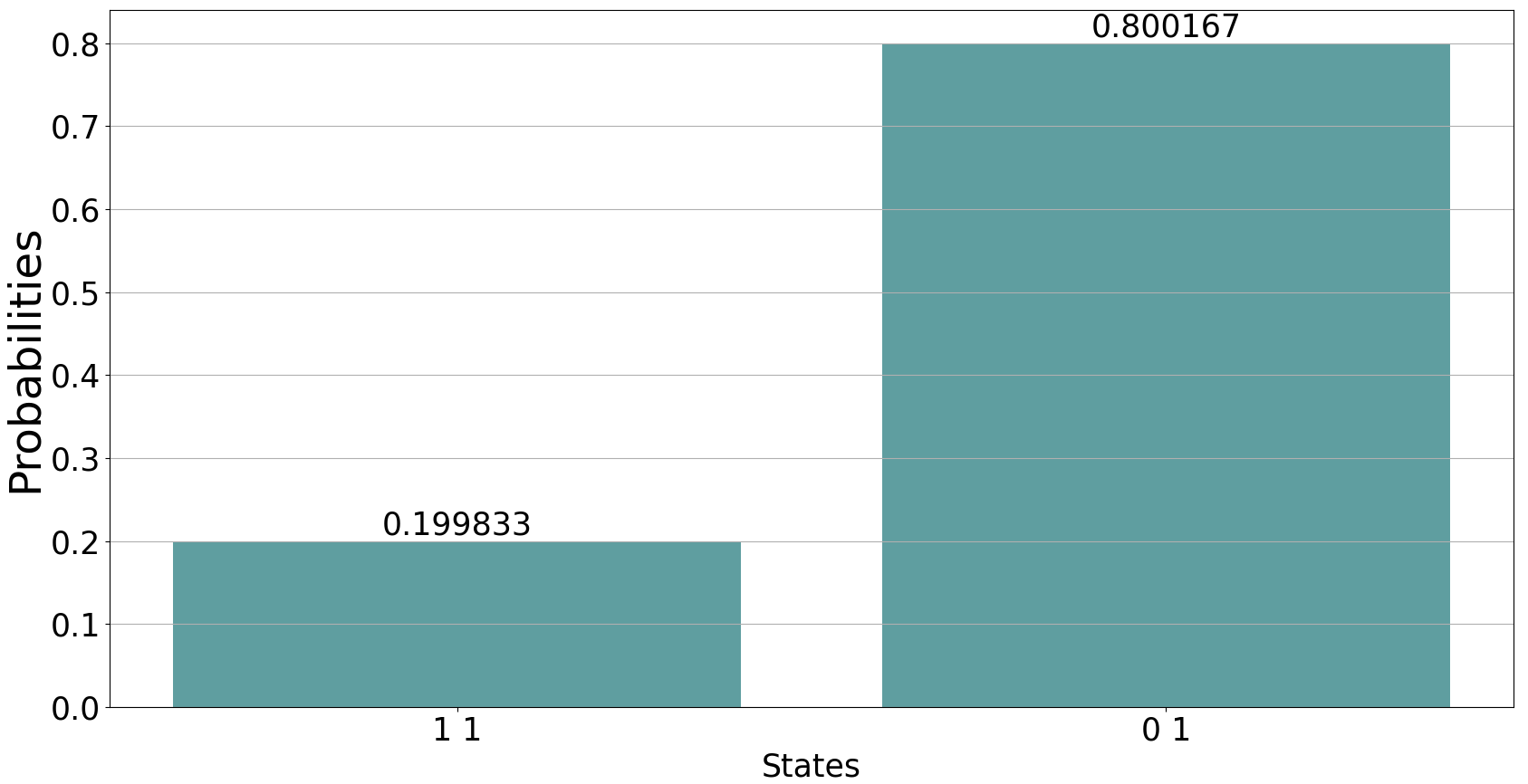}
        \caption{Average Probabilities}
        \label{fig:exact-encoding-probs}
    \end{minipage}
    \hfill
    \begin{minipage}[b]{0.49\textwidth}
        \includegraphics[width=\textwidth]{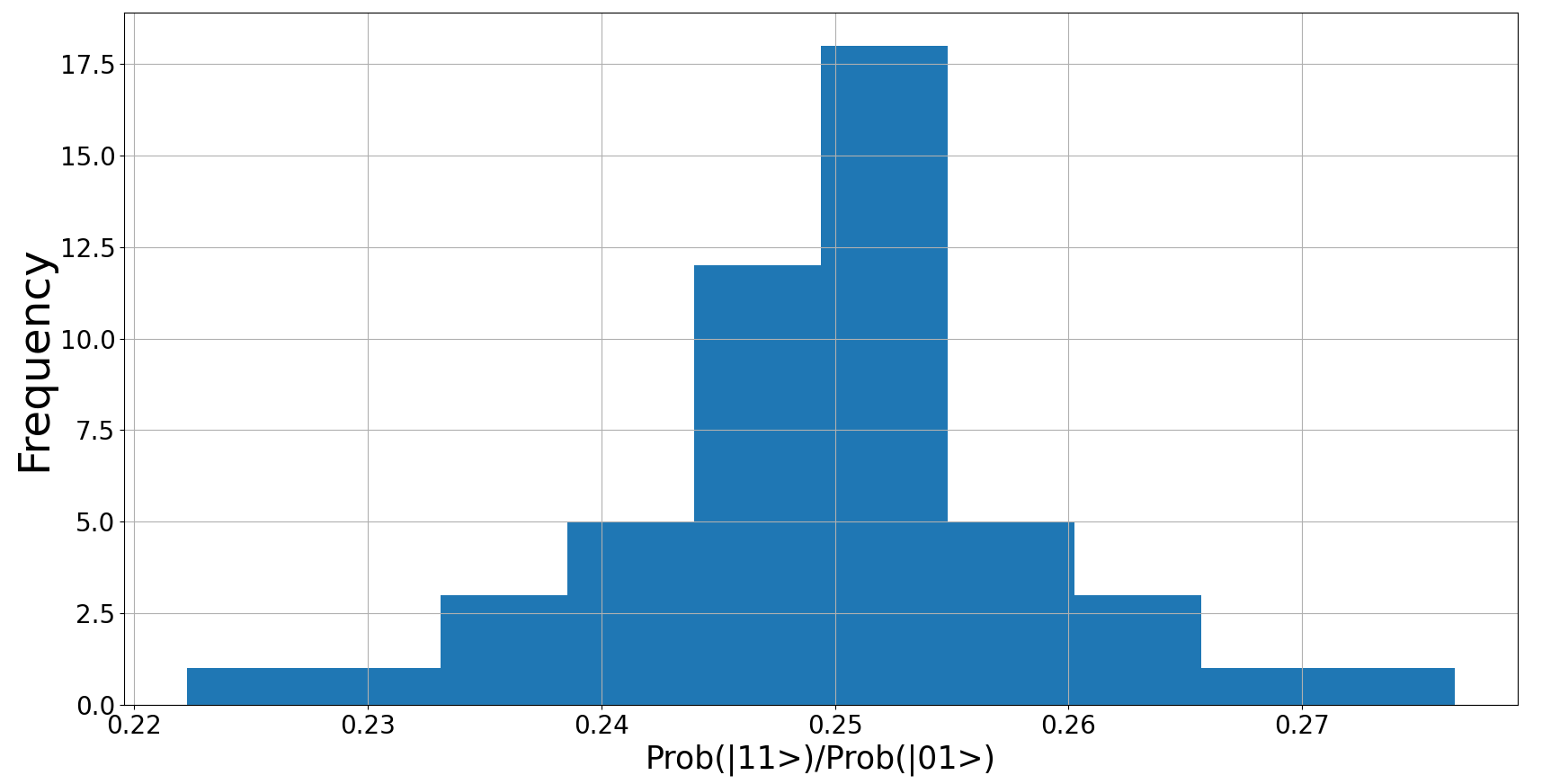} 
        \caption{Distribution of Ratio of Probabilities}
        \label{fig:exact-encoding-ratio}
    \end{minipage}
\end{figure}

As an example, consider the problem,
\begin{equation}A=
   A=\begin{pmatrix}
        1&-1/2\\-1/2&1.
\end{pmatrix}\quad\text{and}\quad\vec{b}=\begin{pmatrix}
        1\\0
    \end{pmatrix}.
\end{equation}

which has the solution  $\vec x=\frac{1}{3}(4,2)$. As such, one expects the ratio of probabilities of the result states to be $(4/3)^2:(2/3)^2=4:1$. We simulate this system $50$ times with each simulation amounting to $10^4$. The result of these simulations is presented in Figures \ref{fig:exact-encoding-probs} and \ref{fig:exact-encoding-ratio}. We find that the ratio of probabilities does match with the expected value of $4:1$. 

\section{Optimisation}


As mentioned above, the HHL algorithm relies upon successfully measuring $\ket 1$ on the ancilla qubit. The form of $A$, and in particular, its condition number $\kappa$ have a significant impact on the probability of this outcome. Indeed, for most interesting problems the problem matrix has high condition number, which decreases this probability. Several ways to circumvent this limitation have been discussed in the literature and they range from modifying $A$ to lower the condition number without altering the problem, to modifying the HHL algorithm itself\cite{Brassard,Ambainis,clader13,Shao,Tong,Babukhin,Childs,calcuttagroup25}.

In this work we discuss results obtained with two optimisation strategies that improve fidelity and scalability on near-term simulators. The dominant cost in HHL arises from the repeated applications of controlled $e^{iAt}$ operations in QPE. Consequently, our approaches focus on simplifying Hamiltonian simulation so that these operators can be realised optimally.

\subsection{Trotterisation}

\begin{figure}[h]
                \centering
                \includegraphics[width=0.85\textwidth]{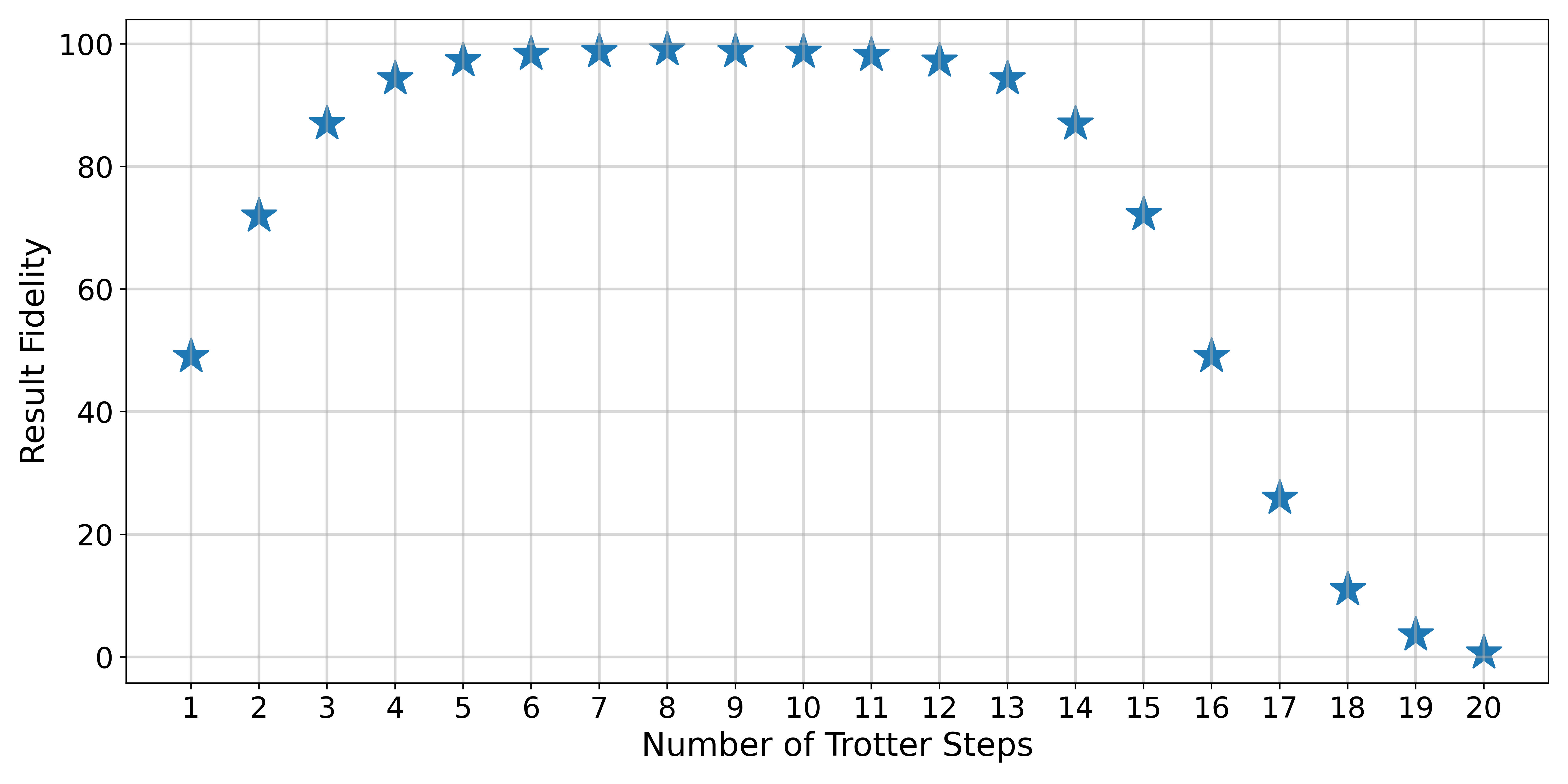}
                \caption{Effect of number of Trotter Steps on Result Fidelity for a Tridiagonal Matrix}
                \label{fig:trottersteps}
\end{figure}

The first of the two methods we discuss is the use of a Suzuki-Trotter decomposition for $e^{iAt}$ and instead implement it in QPE. In practice, lower-order Lie–Trotter–Suzuki product formulas \cite{Ostmeyer_2023},
\begin{equation}
    \mathrm{e}^{h \sum_{k=1}^{\Lambda} A_k+\mathcal{O}\left(h^{n+1}\right)}\approx\left(\prod_{k=1}^{\Lambda} \mathrm{e}^{A_k c_1 h}\right)\left(\prod_{k=\Lambda}^1 \mathrm{e}^{A_k d_1 h}\right) \cdots\left(\prod_{k=1}^{\Lambda} \mathrm{e}^{A_k c_q h}\right)\left(\prod_{k=\Lambda}^1 \mathrm{e}^{A_k d_q h}\right),
\end{equation}

provided the most favourable depth–accuracy trade-off. While increasing the number of Trotter steps reduced short-time simulation error, overall fidelity eventually degraded due to accumulated gate operations and control overhead in the controlled operations. This is presented in Figure \ref{fig:trottersteps}. For matrices of moderate sparsity, we observed an optimal intermediate regime in which simulation accuracy and circuit depth were balanced most effectively. These findings are consistent with the expectation that asymptotically efficient product formulas may be impractical on depth-limited platforms.

 \begin{figure}[h]
                \centering
                \includegraphics[width=0.8\textwidth]{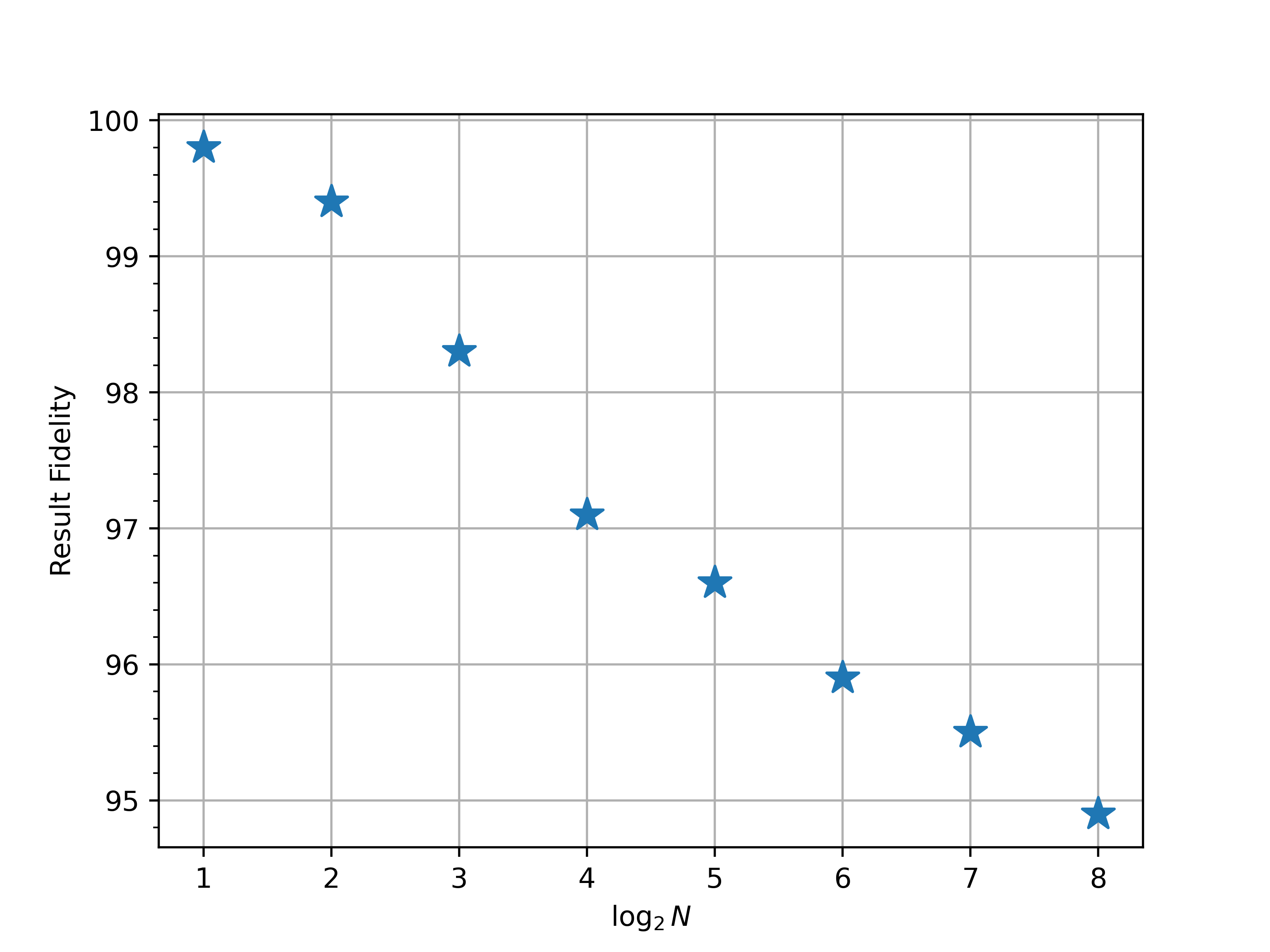}
                \caption{Result Fidelities for HHL with Trotterisation}
                \label{fig:trotter}
\end{figure}

The result fidelities $|\braket{\psi_{exact}|\psi_{HHL}}|^2\times100$ as a functions of the system size for this method are presented in Figure \ref{fig:trotter}.

The fidelity decreases with system size but looking closely there appears to be a decrease in the slope and the hint of some plateau as we increase the system size. We intend to explore this further and ultimately produce a properly scalable implementation of the HHL algorithm.

\subsection{Block encoding}

In our second method, we use block encoding \cite{blockencoding} to embed the matrix $A$ into a larger unitary acting on an extended Hilbert space,

\begin{equation}
    U =
\begin{pmatrix}
A & \cdot \\
\cdot & \cdot
\end{pmatrix},
\end{equation}

which allows one to directly implement $A$ and use $e^{iAt}=\sum_j(iAt)^j/j!$ to implement the controlled unitaries. This method has been seen to perform better when the matrix $A$ is simple but $e^{iAt}$ is harder to implement compared to Trotterisation for reasons discussed below.

Block encoding reduces simulation error per controlled operation and consistently achieves higher result fidelity than Trotterisation for comparable logical precision, particularly for moderately dense matrices. The principal limitation of this approach is that is requires the use of additional ancillary qubits to extend the Hilbert space, which restricts accessible system sizes on qubit-constrained simulators and near-term devices.

\begin{figure}[h]
                \centering
                \includegraphics[width=0.8\textwidth]{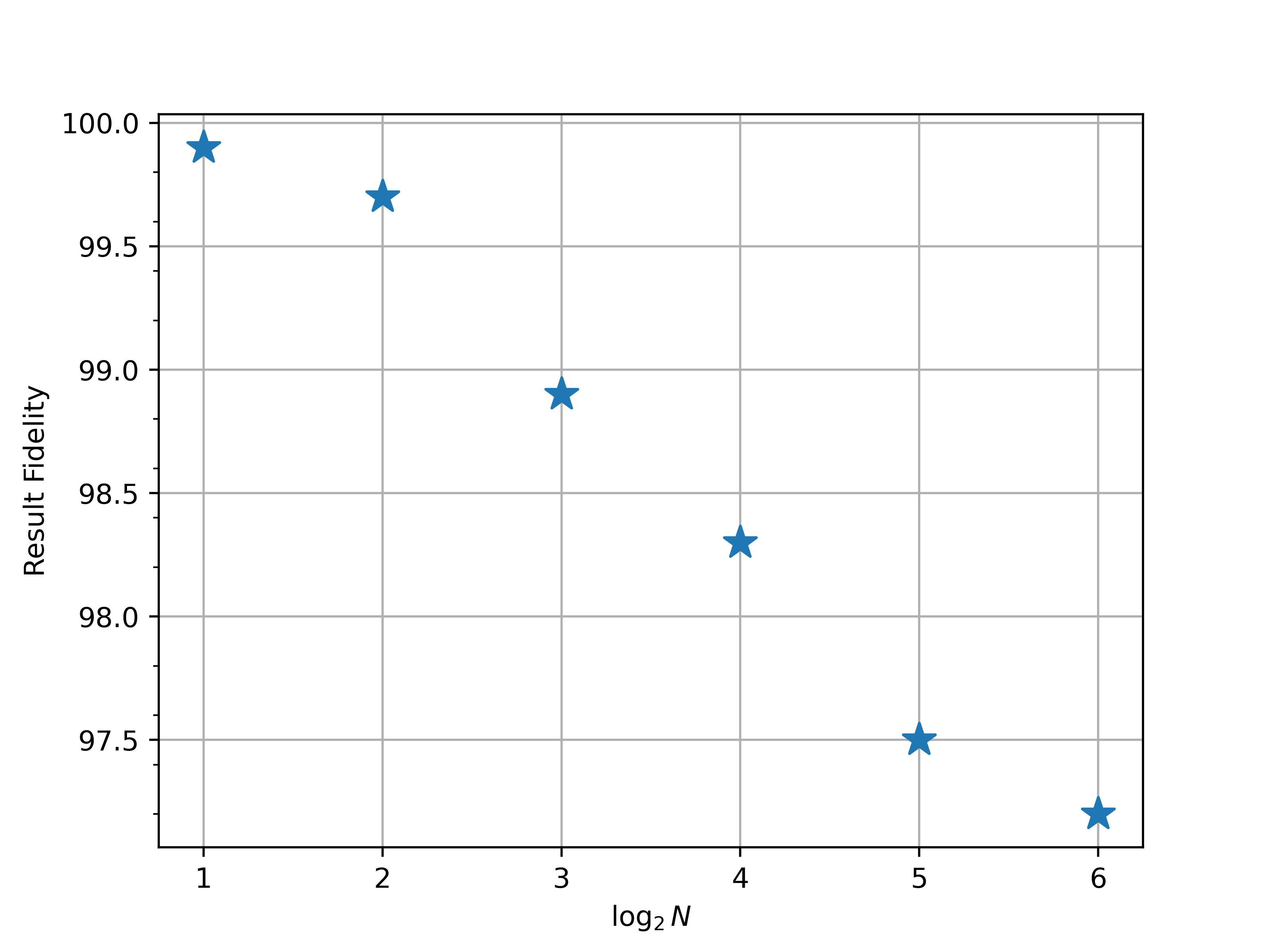}
                \caption{Result Fidelities with Block Encoding}
                \label{fig:encoding}
        \end{figure}

A similar plot of the result fidelities for block encoding is given in Figure \ref{fig:encoding}. We see similar monotonously decreasing behaviour as with Trotterization. However at larger $N,$ the rate of decrease of fidelity is slower.
        
\section{Results}

\begin{description}

    \item[Diagonal Matrices] Although a trivial case, consider first the example of diagonal matrices. Here the Hamiltonian simulation is exact up to numerical precision, and QPE incurs minimal approximation error. In this regime the dominant limitation is the availability of qubits rather than circuit depth. We successfully simulated systems up to $N=1024$ with an average fidelity of approximately $99.3\%$, demonstrating near-ideal behaviour. This case serves as a useful baseline, illustrating that when eigenvectors coincide with the computational basis, HHL performs essentially optimally.

    \item[Tridiagonal Matrices] Tridiagonal matrices retain high sparsity while introducing nontrivial Hamiltonian simulation. Using optimised low-order Trotterisation, systems up to $N=256$ were solved with an average fidelity of $94.9\%$. The observed fidelity loss is consistent with accumulated Trotter error and finite QPE precision, rather than imperfections in state preparation. These results highlight that moderate sparsity still permits favourable scaling, provided simulation depth is carefully controlled.

    \item[Moderately Dense Matrices] For matrices with fewer than $N/2$ nonzero entries per row, achievable system sizes decreased substantially. We report results up to $N=32$ with an average fidelity of $91.7\%$. In this regime, block encoding consistently outperformed Trotterisation for fixed logical precision, reflecting its reduced simulation error per controlled operation. However, the additional ancilla qubits required by block encoding constrained further scaling.

    \item[Fully Dense Matrices] Dense matrices proved the most challenging case. For $N=16$, average fidelity was $90.5\%$, dropping to $80.3\%$ at $N=32$. This rapid degradation reflects the combined impact of increased Hamiltonian simulation cost, deeper QPE circuits, and reduced post-selection success probability associated with small eigenvalues. These results underscore the practical necessity of preconditioning or hybrid classical–quantum strategies when applying HHL to dense systems.

    \item[Interpretation and Scaling Trends] Across all cases, matrix structure and sparsity emerged as the dominant factors governing practical HHL performance. High sparsity enables excellent fidelity and favourable scaling, while dense matrices quickly exceed feasible simulation resources. These trends are consistent with theoretical expectations for Hamiltonian simulation complexity and reinforce the view that HHL is best suited to well-structured linear systems.

\end{description}

\section{Conclusion and Outlooks}

In this work we investigated the practical implementation and optimisation of the Harrow–Hassidim–Lloyd (HHL) algorithm, with particular emphasis on Hamiltonian simulation strategies and empirical performance on simulators. Our results confirm that, while HHL offers favourable asymptotic scaling in the system dimension, its practical performance is governed primarily by matrix structure, sparsity, and spectral properties. For highly structured cases such as diagonal matrices, HHL achieves near-ideal fidelities even at large system sizes, whereas decreasing sparsity leads to rapidly increasing simulation cost and a corresponding loss of fidelity. These trends are fully consistent with theoretical expectations for Hamiltonian simulation and quantum phase estimation (QPE).


Overall, our findings indicate that HHL is best suited to well-conditioned, structured linear systems where problem-specific features can be exploited. 
Future work will focus on hybrid classical–quantum pipelines incorporating preconditioning, low-depth QPE with classical refinement, hardware-aware compilation, and error mitigation strategies. This will permit implementation on quantum hardware to translate the theoretical advantages of HHL into practical gains on near- and mid-term quantum devices.

\acknowledgments
This work is supported by the Department of Atomic Energy, Government of India, under Project Identification Number RTI-4012.
Computations were carried out on the computing clusters at the Department of Theoretical Physics, TIFR, Mumbai. We would also like to thank  Ajay Salve and Kapil Ghadiali for computational support.  We acknowledge the use of Qskit@IBM \cite{qiskit2024} and Cuquantum@Nvidia software \cite{the_cuquantum_development_team_2023_10068206} for quantum simulations.

\bibliographystyle{JHEP}
\bibliography{HHL}



\end{document}